\documentclass[apj]{emulateapj}
\usepackage{graphicx}

\begin{document}
\title{Active Galactic Nuclei Selected from {\em GALEX\/}
Spectroscopy: The Ionizing Source Spectrum at $z\sim 1$\altaffilmark{1,2}}
\author{
Amy~J.~Barger,$\!$\altaffilmark{3,4,5} \email{barger@astro.wisc.edu}
Lennox~L.~Cowie$\!$\altaffilmark{5} \email{cowie@ifa.hawaii.edu}
}
\altaffiltext{1}
{Based in part on data obtained from the Multimission
Archive at the Space Telescope Science Institute (MAST).  
STScI is operated by the Association of Universities for 
Research in Astronomy, Inc., under NASA contract NAS5-26555.  
Support for MAST for non-{\em HST\/} data is provided by the 
NASA Office of Space Science via grant NAG5-7584 and by other 
grants and contracts.}
\altaffiltext{2}
{Based in part on data obtained at the W. M. Keck
Observatory, which is operated as a scientific partnership among
the California Institute of Technology, the University of
California, and NASA and was made possible by the generous financial
support of the W. M. Keck Foundation.}
\altaffiltext{3}{
Department of Astronomy, University of
Wisconsin-Madison, 475 North Charter Street, Madison, WI 53706.}
\altaffiltext{4}{
Department of Physics and Astronomy,
University of Hawaii, 2505 Correa Road, Honolulu, HI 96822.}
\altaffiltext{5}{
Institute for Astronomy, University of Hawaii,
2680 Woodlawn Drive, Honolulu, HI 96822.}

\shorttitle{The Ionizing Source Spectrum at $z\sim1$}
\shortauthors{Barger \& Cowie\/}

\slugcomment{In press at The Astrophysical Journal}

%------------------------------
%   Abstract
%------------------------------
\begin{abstract}
We use a complete sample of Ly$\alpha$ emission-line 
selected active galactic nuclei (AGNs) obtained from 
nine deep blank fields observed with the grism 
spectrographs on the {\em Galaxy Evolution Explorer 
(GALEX)\/} satellite to measure the normalization and spectral
shape of the AGN contribution to the ionizing background 
(rest-frame wavelengths 700-900~\AA) at $z\sim 1$.
Our sample consists of 139 sources selected 
in the redshift range $z=0.65-1.25$ in the near-ultraviolet 
(NUV; 2371~\AA\ central wavelength) channel. 
The area covered is 8.2~deg$^2$ to a NUV 
magnitude of 20.5 (AB) and 0.92~deg$^2$ at the faintest  
magnitude limit of 21.8.  The {\em GALEX\/} AGN luminosity 
function agrees well with those obtained using optical and 
X-ray AGN samples, and the measured redshift evolution
of the ionizing volume emissivity is similar to that 
previously obtained by measuring the {\em GALEX\/} 
far-ultraviolet (FUV; 1528~\AA\ central wavelength) magnitudes 
of an X-ray selected sample.
For the first time we are able to construct the shape of the 
ionizing background at $z\sim1$ in a fully self-consistent way.
\end{abstract}


%--------------------------------------------------------
%    1. Introduction
%--------------------------------------------------------
\section{Introduction}
\label{secintro}

An important parameter in the cosmological modeling 
of galaxies and the intergalactic gas is the ionizing
background radiation produced by the overall population 
of active galactic nuclei (AGNs) and star-forming 
galaxies as a function of wavelength and redshift.
Usually the input ionizing source spectrum for such models
is computed by convolving an approximate quasar spectral 
energy distribution (SED) determined from composite quasar 
spectra with a quasar luminosity function determined 
from optical data (e.g., Haardt \& Madau 1996; 
Madau et al.\ 1999; Haardt \& Madau 2001; Meiksin 2005; 
Siana et al.\ 2008; Dall'aglio et al.\ 2009).
(Note that Gilmore et al.\ 2009 
adopted the Hopkins et al.\ 2007 quasar luminosity function, 
which tries to take into account obscured sources, but 
obscured sources will not contribute to the ionizing 
background.)  However, using
optical quasar luminosity functions that have been extrapolated 
beyond where they have been measured (especially if the 
turn-over has not been observed) can result 
in overestimates of the ionizing background.
X-ray luminosity functions, which probe to
fainter luminosities (e.g., Barger et al.\ 2005; 
Richards et al.\ 2005), provide a better approach
(Cowie et al.\ 2009; hereafter, CBT09).

In CBT09 we 
measured the ionizing fluxes from a wide-field (0.9~deg$^2$) 
X-ray sample with optical and UV imaging observations
and found that the AGN contribution to the ionizing 
background peaks at $z\sim2$. These results are
lower than previous estimates of the
AGN contribution, confirming that ionization from AGNs 
is insufficient to maintain the observed ionization of 
the intergalactic medium (IGM) at $z>3$ (see 
Bolton et al.\ 2005 and Meiksin 2005 for earlier estimates 
that suggested this result).

Ideally, however, one would like a
direct measurement of the shape of the input ionizing
source spectrum.  Fortunately, with the advent of the 
{\em Galaxy Evolution Explorer (GALEX)\/} satellite 
(Martin et al.\ 2005), this is now possible at $z\sim 1$.  
Obtaining the shape of this spectrum is the primary 
goal of the present paper.  
We use {\em GALEX\/} grism spectroscopic observations 
to measure the AGN contribution to the ionizing background at 
$z\sim 1$ directly from the UV spectra. 
For this we use the Ly$\alpha$ selected sample from 
Cowie et al.\ (2010; hereafter, CBH10), which should 
provide an essentially complete sample of broad-line 
AGNs (BLAGNs) and a nearly complete sample of all 
AGNs contributing to the ionizing flux
in the redshift interval $z=0.65-1.25$. 
We then go a step further and measure the ionizing source 
spectrum at rest-frame wavelengths $700-900$~\AA.
We note, however, that this wavelength coverage is insufficient to 
characterize the full Lyman continuum background.

The outline of the paper is as follows.
In Section~\ref{secgalex} we briefly describe 
the 139 $z=0.65-1.25$ sources in the CBH10 sample. 
In Section~\ref{seccomplete} we test the completeness
of the sample by comparing with an 
X-ray selected AGN sample in the same redshift interval.
In Section~\ref{secLF} we construct the rest-frame UV 
(1450~\AA) luminosity function and compare it 
with the UV luminosity functions derived from X-ray and 
optically selected samples in the same redshift interval. 
In Section~\ref{secion} we determine the ionizing source spectrum
at $z\sim 1$ and show that the normalization
agrees well with recent determinations at this redshift.
In Section~\ref{secshape} we compute the shape of 
the comoving volume emissivity (after correcting for
incompleteness and for the effects of Lyman scattering
and absorption in the intervening intergalactic medium)
versus rest-frame wavelength for the redshift intervals
$z=0.65-0.95$ and $z=0.95-1.25$.
In Section~\ref{secsum} we summarize our results. 

We use a standard $H_{0}$ = 70~km~s$^{-1}$~Mpc$^{-1}$, 
$\Omega_{m}$ = 0.3, and $\Omega_{\Lambda}$ = 0.7 
cosmology throughout. All magnitudes are given in the
AB magnitude system, where an AB magnitude is defined by
$m_{AB}=-2.5\log f_\nu - 48.60$. Here $f_\nu$ is the
flux of the source in units of 
erg~cm$^{-2}$~s$^{-1}$~Hz$^{-1}$

%--------------------------------------------------------
% 2. The GALEX Sample
%--------------------------------------------------------
\section{The {\em GALEX\/} Sample}
\label{secgalex}

CBH10 constructed a sample of 139 Ly$\alpha$ 
emission line-selected sources in the redshift interval 
$z=0.65-1.25$ from nine blank high galactic 
latitude fields with the deepest {\em GALEX\/} grism 
observations. They took the near-UV (NUV) and 
far-UV (FUV) extracted spectra and the NUV 
magnitudes from the Multimission 
Archive at STScI (MAST). Morrissey et al.\ (2007) detail
the spectral extraction techniques used by the {\em GALEX\/}
team in analyzing the grism data and discuss the UV spectra 
and magnitudes. We summarize the fields in 
Table~\ref{tab1}, where we give in column~(1) the {\em GALEX\/} 
name, in columns~(2) and (3) the J2000 right ascension and declination, 
in column~(4) the exposure time in kiloseconds, in column~(5) the limiting 
NUV magnitude to which the spectra were extracted, in column~(6)
the galactic $E(B-V)$ in the direction of the field from Schlegel et al.\ (1998),
in column~(7) the galactic latitude, in column~(8) the number of sources 
with spectra lying within a radius of $32\farcm5$ from the field center
(chosen because in the outermost regions of the fields
there is a higher fraction of poor quality spectra),
and in column~(9) the number of these sources found
to have a Ly$\alpha$ emission line in the redshift range
$z=0.65-1.25$ based on their NUV spectra.
The extracted sources per field constitute nearly complete
samples to the NUV limiting magnitudes given in the table
(i.e., these are NUV selected samples and comprise both 
galaxies and a significant number of stars).
The FUV spectra cover a wavelength range of
$\sim 1300-1800$~\AA\ at a resolution of $\sim 10$~\AA, 
and the NUV spectra cover a wavelength range of 
$\sim 1850-3000$~\AA\ at a resolution of $\sim25$~\AA.

Each of the spectra was corrected for the Milky Way extinction 
in the field using the Schlegel et al.\ (1998) extinction and the 
Cardelli et al.\ (1989) reddening law. 
Because the Galactic extinction
towards these fields is very low (Table~\ref{tab1}), the net effect
is extremely small. On average the shape of the spectra
is nearly unchanged and the normalization is raised by
a factor of 1.08. The low extinctions
in the present sample represent a considerable advantage
over random quasar samples where the extinction for many
of the objects will be high.

CBH10 determined the observed area as a function
of NUV magnitude for each field by computing the ratio of
sources with {\em GALEX\/} spectra
at a given NUV magnitude to sources with that NUV magnitude
in the continuum catalog.
They multiplied this ratio for each field by the area
corresponding to the $32\farcm5$ selection radius and 
summed the results for the 9 fields to form an area-magnitude
relation (their Figure~4(c)). The area is the 8.2~deg$^2$
area of the 9 fields for magnitudes brighter than NUV$=20.5$
and then drops as the limiting magnitudes of each of the 
individual fields is reached. It falls to zero at $\sim 21.8$,
which is the limiting depth of the deepest (GROTH~00) field.

CBH10 describe their search procedure for finding emission lines 
in the {\em GALEX\/} spectra.  Because the 
spectra become very noisy at the edges of the wavelength 
ranges, they restricted their higher redshift search
to the redshift interval $z=0.65-1.25$ corresponding 
to the wavelength range $2006-2736$~\AA\ in the NUV spectrum
for the Ly$\alpha$ emission line.
They measured redshifts for each source with an emission 
line in this wavelength range and split the sources 
into either an AGN class (if there were high-excitation 
lines; usually OVI, NV, or CIV would be present in addition to 
the Ly$\alpha$ line) or a candidate Galaxy class (if there 
were only a single line visible; in this case the line 
was assumed to be Ly$\alpha$ for determining a redshift, 
which they showed was an extremely reliable assumption).
CBH10 found that nearly all of the more 
luminous sources are AGNs (their Figures~4(a) and 4(b)), 
leaving relatively few candidate Galaxies in the redshift 
interval $z=0.65-1.25$ (see also Deharveng et al.\ 2008).
Of these, many were detected in the FUV (their Figure~4(b)), 
suggesting that they are AGNs, too,
since $z\sim1$ galaxies should be very faint or undetected 
in the FUV band, which lies below the Lyman continuum edge 
at these redshifts (e.g., Siana et al.\ 2007; CBT09).
We shall hereafter refer to all of the sources that were 
not classified as AGNs as unclassified sources. Since
nearly all of them are likely AGNs, we will refer to
the combined AGN plus unclassified source sample as our 
``full'' AGN sample.  We note that we have confirmed
four of the unclassified sources as star formers through
optical spectroscopic observations, but excluding them from 
our analysis makes no difference to the results.

%-------------
% TABLE 1
%-------------
\hskip -1.5cm
\begin{deluxetable*}{ccccccccc}
\renewcommand\baselinestretch{1.0}
\tablecaption{{\em GALEX\/} FIELDS}
\tablehead{Name & R.A. & Decl. & Exp. & NUV$_{lim}$ & 
$E(B-V)$ & Gal. Lat. & No. w/sp. & No. w/Ly$\alpha$ \\
& (J2000.0) & (J2000.0) & (ks) & (mag) & & (deg) & ($<32\farcm5$) 
& %($z=0.65-1.25$) 
\\ 
(1) & (2) & (3) & (4)  & (5) & (6) & (7) & (8) & (9)
}
\startdata
GROTH 00  &  214.99182 & 52.78173  &  291  &  21.8  &  0.007  &  59.5 & 1152  &  61  \cr
NGPDWS 00  &  219.15610	& 35.17135   &  156  &  21.5  &  0.009 & 66.2  & 738  &  39  \cr
CDFS 00  &  53.12779  &	-27.87137  &  149  &  21.5  &  0.008  & -54.4 & 876  &  36  \cr
COSMOS 00  &  150.11900	& 2.20583 &  140  &  21.5  &  0.018 & 42.1  & 779  &  27  \cr
ELAISS1 00  &  9.63857	& -43.99023  &  84  &  21.2  &  0.007 & -72.9  & 589  &  17  \cr
SIRTFFL 00  &  259.12387 & 59.90915  &  80  &  21.2  &  0.020 & 35.0  & 800  &  21  \cr
LOCK 00  &  162.67843 &	58.73117  &  48  &  20.9  &  0.009 & 52.1  & 390  &  10  \cr
SIRTFFL 01  &  260.41425 & 59.34286  &  34  &  20.7  &  0.029 & 34.4  & 612  &  14  \cr
HDFN 00  &  189.20946 &	62.19772  &  24  &  20.5  &  0.012 & 54.8  & 265 &  10  \cr
\enddata
\label{tab1}
\end{deluxetable*}

%---------------------------
%   3.  Completeness
%---------------------------
\section{Testing the Completeness of the {\em GALEX\/} AGN Sample}
\label{seccomplete}

\begin{figure}
%3.4 ---> 3.6
\centerline{\includegraphics[width=3.6in,angle=0,scale=1.]{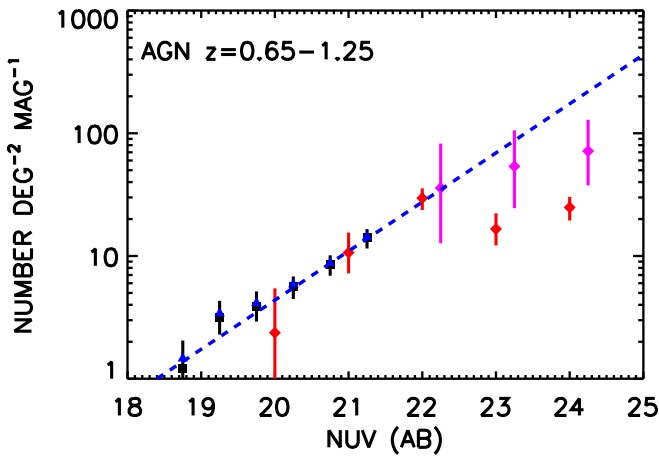}}
\caption{
Number counts of the {\em GALEX\/}
AGNs in the redshift interval $z=0.65-1.25$ vs. NUV 
magnitude (black squares with $1\sigma$ errors based on 
the number of sources in each bin). The blue triangles 
show the values if we also include the unclassified
sources in this redshift interval.  For comparison, we 
show the number counts of an X-ray selected sample in a
0.9~deg$^2$ area (the CLANS and CLASXS fields)
with red diamonds and in a smaller but much deeper 
200~arcmin$^2$ area (the CDF-N) with purple diamonds. 
Over the magnitude range of the {\em GALEX\/} grism data
the {\em GALEX\/} AGN number counts agree well with 
the X-ray selected AGN number counts and are well fit with a 
slope of 0.4 (blue dashed line). Thus, each magnitude bin 
contains roughly equal light density. 
\label{nmag_counts}
}
\end{figure}

We test the completeness of the {\em GALEX\/}
AGN sample by comparing with an X-ray selected 
AGN sample in the same redshift interval. 
In Figure~\ref{nmag_counts} we show the number counts
for the {\em GALEX\/} AGN sample (black squares) and for
the {\em GALEX\/} full AGN sample
(blue triangles) versus NUV magnitude in the 
redshift interval $z=0.65-1.25$. (There is only a very 
small difference between the AGN and the full AGN results.)  
In each magnitude bin the counts are the sum of the 
inverse areas for all the sources in that bin.
The error bars are $\pm~1\sigma$ based on the Poisson 
errors corresponding to the number of sources
in each bin. The number counts per square degree
per unit magnitude, $N$, are well described by the form
(blue dashed line)
\begin{equation}
\log N  = 0.40\ ({\rm NUV} - 18.4)
\label{eqnnuv1}
\end{equation}
to the limiting NUV magnitude of the grism data of 21.8.
This functional form implies an equal 
amount of light in each magnitude interval.

%--------------------------------------------------------------------------------------
% FIGURE 1.  NUMBER COUNTS OF BLAGN:  nmag_counts.pro
%--------------------------------------------------------------------------------------
%--------------------------------------------------------------

For the X-ray selected AGN sample in the same 
redshift interval we use the extremely spectroscopically 
complete {\em Chandra\/} samples of 
Trouille et al.\ (2008, 2009) and references therein
on the {\em Chandra\/} Large-Area Synopotic X-ray Survey 
(CLASXS), the {\em Chandra\/} Lockman Area North Survey (CLANS), 
and the {\em Chandra\/} Deep Field-North (CDF-N). 
Essentially all of the X-ray sources in the
redshift interval $z=0.65-1.25$ are expected 
to be identified. 

The CLASXS and CLANS fields are covered by a number of
{\em GALEX\/} pointings, all of which are roughly 30~ks
in depth. The CDF-N {\em GALEX\/} exposures are 
165~ks in the NUV and 96~ks in the FUV.  CBT09 describe 
how the {\em GALEX\/} magnitudes for the sources in the 
X-ray samples were measured. In brief, given the large point-spread
function (PSF) of {\em GALEX\/} ($4\farcs5-6\farcs0$ FWHM),
they used an $8''$ diameter aperture to measure the magnitudes
using the {\em GALEX\/} zeropoints from Morrissey et al.\ (2007),
and they corrected these to approximate total magnitudes using
an offset that they determined to be $-0.41$. They measured the noise 
level in the images by randomly positioning apertures on blank 
regions of sky and measuring the dispersion. In the CLASXS and
CLANS fields they found the $1\sigma$ limits to be slightly 
variable but typically about 26.1 in the NUV images 
and 26.5 in the FUV images. In the CDF-N field they found 
$1\sigma$ limits of 26.8 in the NUV image and 27.4 in the FUV 
image. Only a small fraction ($\sim 4$\%) of the X-ray sources 
in the CLASXS or CLANS fields are not covered by one of the 
{\em GALEX\/} pointings, and all of the CDF-N sources are 
covered. CBT09 only eliminated four sources where visual 
inspection showed that there was contamination from a nearby 
brighter {\em GALEX\/} source at the same wavelength.
The X-ray sources themselves are so sparse that contamination
by another X-ray source may be neglected.

In Figure~\ref{nmag_counts} we show separately the number 
counts from the shallower CLASXS and CLANS fields
(red diamonds), where the $2-8$~keV flux limit is 
$\sim7\times10^{-15}$~erg~cm$^{-2}$~s$^{-1}$ and
the area is $\sim0.9$~deg$^2$, and the number counts 
from the CDF-N (purple diamonds), which is about 50 times 
deeper but has a much smaller area of $\sim200$~arcmin$^2$. 
Over the magnitude range of the {\em GALEX\/} grism data
the {\em GALEX\/} AGN number counts agree extremely well 
with the CLASXS and CLANS AGN number counts.
When we compare the NUV counts from the shallower
X-ray fields with those from the CDF-N,
we see that the former are becoming incomplete fainter
than NUV~$\sim22.5$.
The CDF-N number counts themselves should be complete 
to several magnitudes fainter than this, so their behavior
suggests that the AGN number counts begin to turn down 
from the 0.4 slope at NUV~$\sim23.5$. 
This in turn suggests that the {\em GALEX\/} AGN 
sample measures a large fraction of the total NUV light 
at these redshifts.

%------------------------------------------------------------------
% FIGURE 2.  UV luminosity function:  lum_fun.pro 
%------------------------------------------------------------------
\begin{figure}
%3.4 ---> 3.6
\hskip 0.4cm
\centerline{\includegraphics[width=3.6in,angle=0,scale=1]{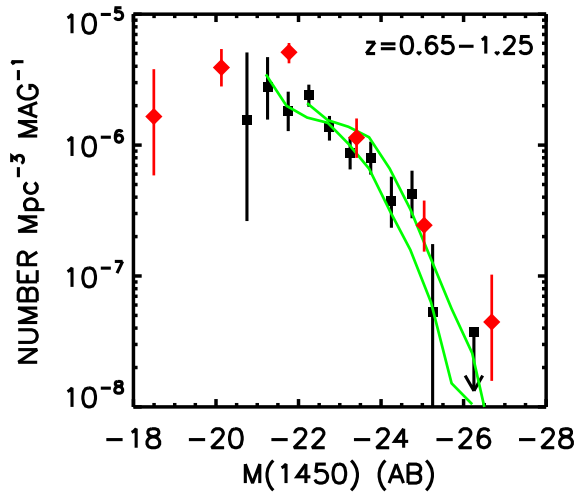}}
\caption{
The rest-frame UV (1450~\AA) luminosity function of the
{\em GALEX\/} full AGN sample in 
the redshift interval $z=0.65-1.25$ (black squares with 
$1\sigma$ error bars from the observed number of sources 
in each bin).
Red diamonds show the converted X-ray selected BLAGN 
luminosity function in the same redshift interval
from Yencho et al.\ (2009).
The green curves (lower for $z=0.68-0.95$; upper for
$z=0.95-1.25$) show the converted optically selected 
luminosity functions from Croom et al.\ (2004).
\label{lum_fun}
}
\end{figure}
%--------------------------------------------------------------

%------------------------
%   4.  The AGN LF
%------------------------
\section{The AGN Luminosity Function}
\label{secLF}
We next computed the binned rest-frame UV (1450~\AA) luminosity 
function for the {\em GALEX\/} full AGN sample using the 
NUV-continuum magnitudes and the $1/V$ method (Felten 1976).
We made the small differential $K-$corrections to 1450~\AA\
by assuming a quasar spectral energy distribution (SED) 
with $f_\nu \sim \nu^{-0.44}$ (Vanden Berk et al.\ 2001). 
We show this luminosity function with black squares 
in Figure~\ref{lum_fun}.
The error bars are $\pm1\sigma$ based on the Poisson errors
corresponding to the number of sources in each bin.  
We compare with the 1450~\AA\ luminosity function 
derived from the X-ray selected BLAGN luminosity function
of Yencho et al.\ (2009) (red diamonds). 
For the conversion of $2-8$~keV luminosity ($L_X$)
to 1450~\AA\ absolute magnitude we used the relation
$\nu L_\nu (2300~{\rm \AA})/L_X=7.46\times 
(L_X/10^{44}~{\rm erg}~{\rm s}^{-1})^{0.31}$ from CBT09,
which is similar to previous work 
(e.g., Vignali et al.\ 2003). 
We again made the small differential $K-$corrections to 
1450~\AA\ by assuming $f_\nu \sim \nu^{-0.44}$.
Finally, we compare with the optically selected 
quasar luminosity function of Croom et al.\ (2004) 
(green curves) scaled to 
1450~\AA\ by assuming $f_\nu \sim \nu^{-0.44}$.
A similar result is obtained using the 
Richards et al.\ (2005) luminosity function. 

The {\em GALEX\/} luminosity function agrees 
extremely well with these other determinations and reaches 
about a magnitude fainter than the wide-field optical
samples used in the Croom et al.\ (2004) measurement.
The X-ray-determined luminosity function from 
Yencho et al.\ (2009) 
is deeper yet and shows a turn-down at fainter magnitudes. 
This means that the UV light density is nearly
convergent in the {\em GALEX\/} full AGN sample at these 
redshifts. If we extrapolate the {\em GALEX\/} 
luminosity function with a constant value below 
$M(1450)_{\rm AB}=-22$, then the additional UV light 
density at magnitudes fainter 
than $M(1450)_{\rm AB}=-22$ is only 24\% of the total 
light for the observed sources with magnitudes brighter 
than this value.

%---------------------------------------------------------------------------------
% FIGURE 3.  ionization evolution:  nmg_ion.pro, nmg_ionl.pro
%---------------------------------------------------------------------------------
\begin{figure}
\centerline{
\includegraphics[width=3.6in,clip,angle=0,scale=1]{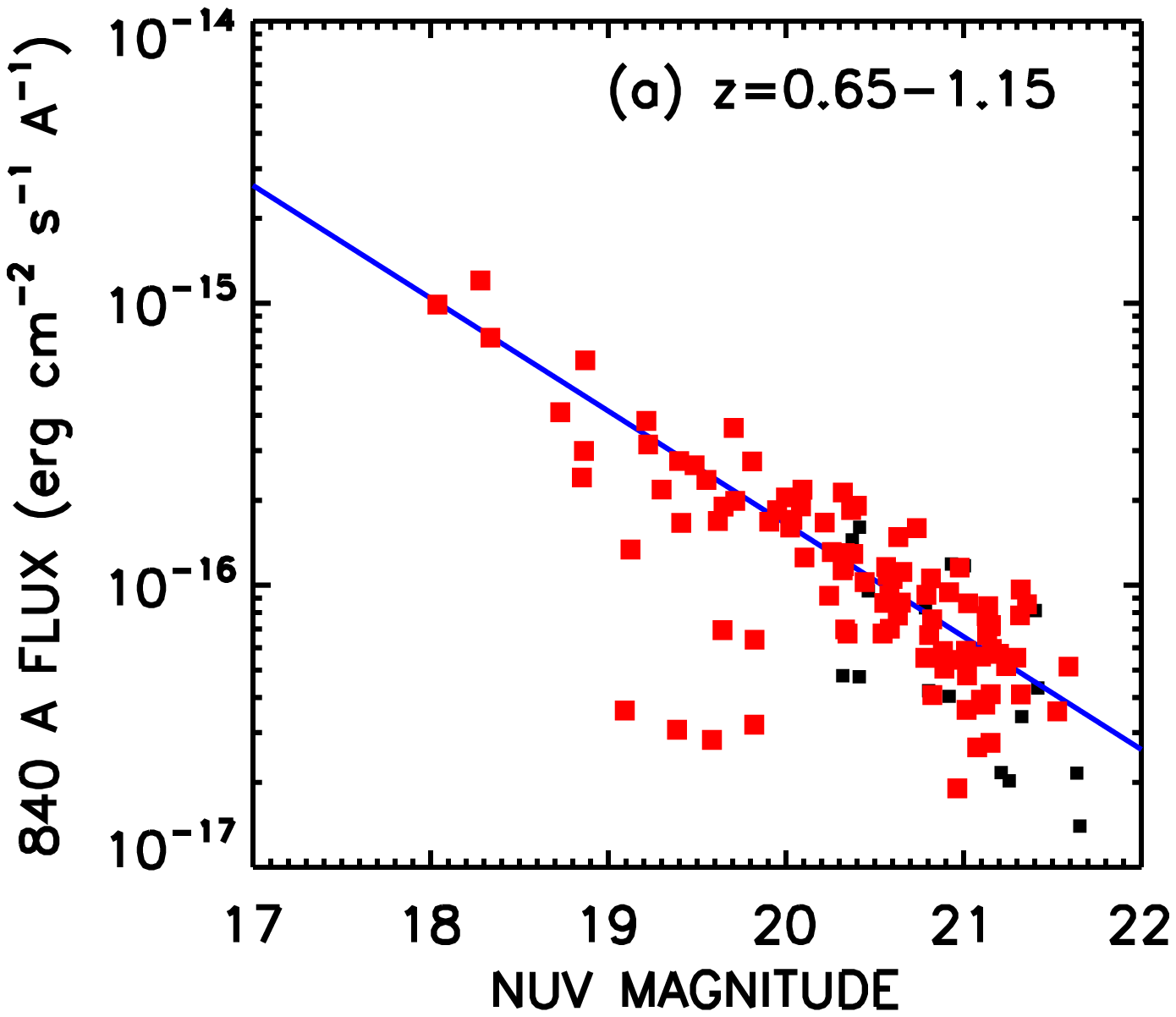}}
\centerline{\includegraphics[width=3.6in,clip,angle=0,scale=1]{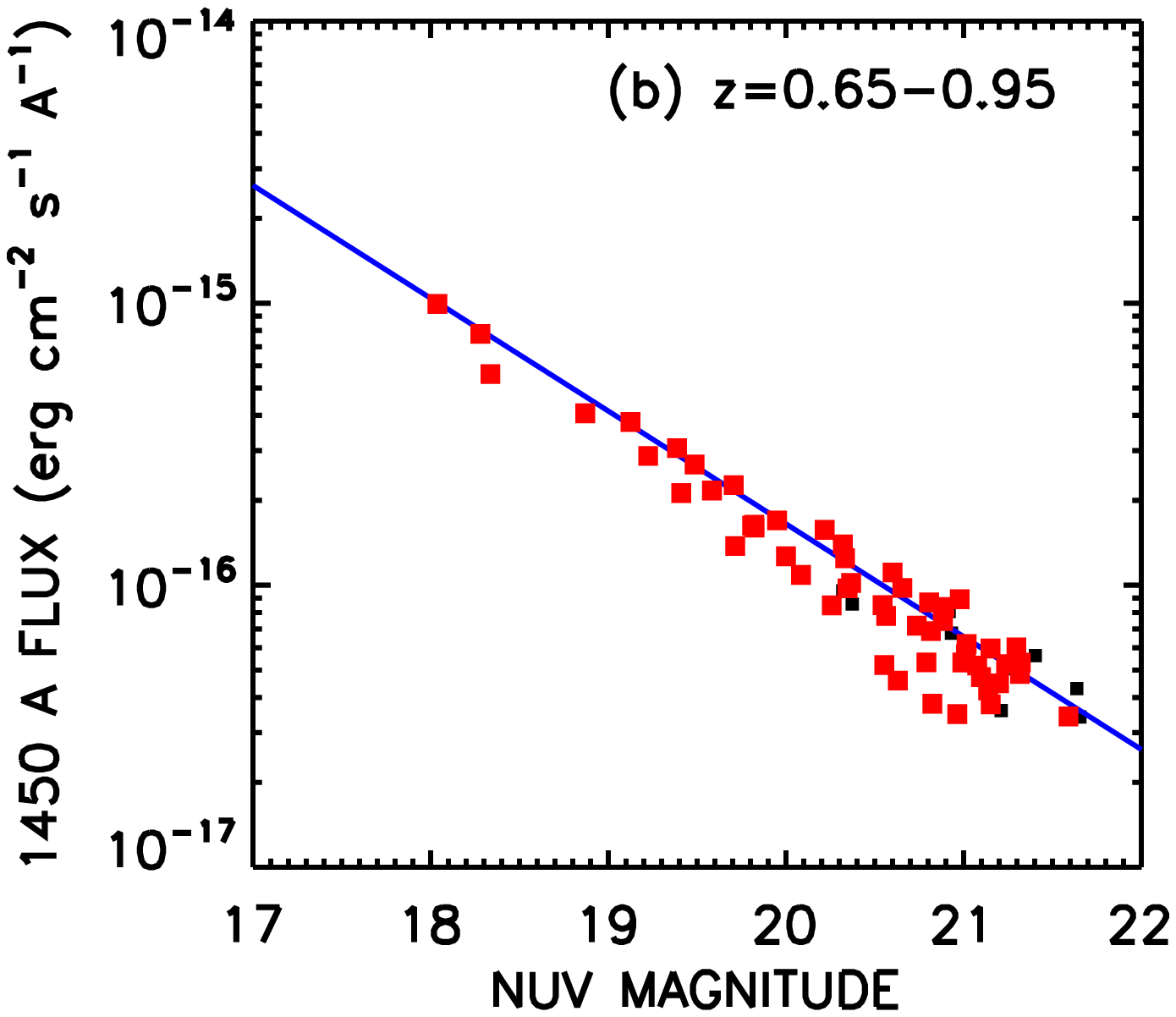}
}
\caption{
(a) Observed flux from the {\em GALEX\/} 
spectra at rest-frame $800-880$~\AA\ vs. NUV magnitude
for sources in the redshift interval $z=0.65-1.15$.
Red squares denote sources classified as AGNs due to the 
presence of high-excitation lines, while black squares 
show the unclassified sources in this redshift interval.  
The blue line shows a linear fit to the data.
(b) Same as (a) but for rest-frame $1400-1500$~\AA\ and
restricted to $z=0.65-0.95$, where the {\em GALEX\/} 
NUV spectra cover the rest-frame wavelengths.
The blue line shows the expected value of the flux
determined from the NUV continuum magnitude, since redshifted
$1400-1500$~\AA\ matches closely to the NUV wavelength.
There is a very small multiplicative factor (0.95)
between the flux measured from the spectra and the 
flux measured from the broadband magnitudes.
\label{nmg_ion}
}
\end{figure}
%--------------------------------------------------------------

%------------------------------------------------------
%   5.  The Ionizing Spectrum of z~1 AGN
%------------------------------------------------------
\section{The Ionizing Spectrum of $z\sim1$ AGNs} 
\label{secion}

We next compared the flux measured from the spectra at rest-frame
wavelengths both above and below the Lyman continuum break 
with the selection NUV magnitude.
In Figure~\ref{nmg_ion} we show the observed
flux from the {\em GALEX\/} spectra at rest-frame
(a) $800-880$~\AA\ (for $z=0.65-1.15$ only; note that we 
use only the portion of the wavelength range that is covered
by the {\em GALEX\/} FUV spectra) and 
(b) $1400-1500$~\AA\ (for $z=0.65-0.95$ only, in order 
for the {\em GALEX\/} NUV spectra to cover the rest-frame
wavelengths) versus NUV. 
The tight relation in (b) simply reflects
the fact that the redshifted $1400-1500$~\AA\
spectral interval matches closely to the NUV wavelength. 
The data provide a test of the {\em GALEX\/} spectral 
intensity calibration, which appears to be extremely accurate, 
since the spectral fluxes differ by less than 5\% from the 
broadband continuum fluxes.

As can be seen from Figure~\ref{nmg_ion}(a), most of the 
sources also obey a linear relation between the spectral 
flux at wavelengths just below the continuum break and the NUV 
flux. At 840~\AA\ we find 
\begin{equation}
f_\lambda(840~{\rm \AA})=1.65\times10^{-16} \ 10^{-0.4 ({\rm NUV}-20)} \,,
\label{eqnnuv}
\end{equation}
where $f_\lambda(840~{\rm \AA})$ is the average spectral 
flux of the sources at rest-frame $800-880$~\AA\
in erg~cm$^{-2}$~s$^{-1}$~\AA$^{-1}$.
We show this as the blue line in Figure~\ref{nmg_ion}(a).
A small fraction of the sources (i.e., those where there 
is a significant continuum break) lie significantly 
below this relation.  The mean $f_\lambda$ ratio,
$f_\lambda(840~{\rm \AA})/f_\lambda(1450~{\rm \AA})$,
is 1.11; the mean $f_{\nu}$ ratio is 0.37.
CBT09 gave an $f_{\nu}$ ratio of 0.26 between 912~\AA\ 
and 2300~\AA, which for $f_\nu\sim\nu^{-0.44}$ 
(Vanden Berk et al.\ 2001) would correspond to 0.34 for 
the $f_{\nu}$(840~\AA)$/f_{\nu}$(1450~\AA) ratio,
in close agreement with the present value.

%-----------------------------------------------------------------------------------------
% FIGURE 4.  cumulative volume emissivity:  ion_vol_cumulative.pro
%-----------------------------------------------------------------------------------------
\begin{figure}
\hskip 0.4cm
\centerline{\includegraphics[width=3.6in,clip,angle=0,scale=1]{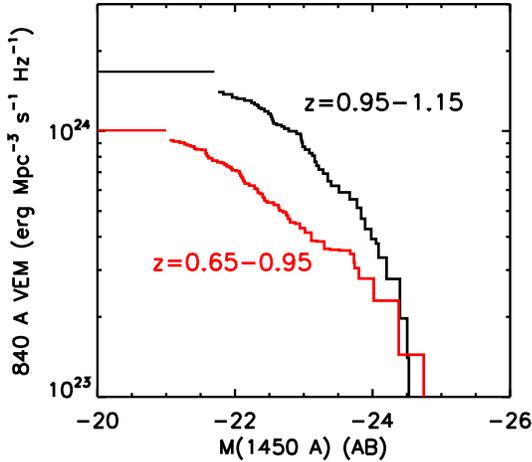}}
\caption{Cumulative comoving volume emissivity (VEM) for the 
full AGN sample at rest-frame 840~\AA\ vs. rest-frame 1450~\AA\ 
absolute magnitude for the two redshift intervals $z=0.65-0.95$ 
(red curve) and $z=0.95-1.15$ (black curve).
The horizontal lines show the incompleteness corrected values
for the two redshift intervals.
\label{vem_cum}
}
\end{figure}
%--------------------------------------------------------------

We then computed the comoving volume emissivity (VEM) in the 
rest-frame wavelength range $800-880$~\AA\ using the $1/V$
methodology.  In Figure~\ref{vem_cum} we show the
cumulative contribution to the VEM
as a function of the rest-frame 1450~\AA\ absolute magnitude 
for the redshift intervals $z=0.65-0.95$ and $z=0.95-1.15$,
which reach $M_{\rm AB}$(1450~\AA)$~\sim -21$ and $\sim -21.7$, 
respectively. 

We have taken a simple approach to estimate 
the incompleteness correction due to sources fainter 
than our selection NUV magnitude limit.  
As a result of the two linear relations in Figure~\ref{nmg_ion},
the $f_\lambda(840~{\rm \AA})/f_\lambda(1450~{\rm \AA})$ ratio 
stays fixed over the NUV magnitude range of our
sources, which implies that the spectral shape
is invariant over that magnitude range. 
If we assume that the ratio also
remains fixed at fainter magnitudes, then any estimate of 
the incompleteness correction made at 1450~\AA\ can be applied 
at any other wavelength.

In order to estimate the incompleteness correction at 1450~\AA,
we used the form of the 1450~\AA\ luminosity function
(e.g., Figure~\ref{lum_fun}).  We determined the amount of 
missing light at 1450~\AA\ by assuming a constant value 
per Mpc$^3$ per mag below the limiting absolute magnitude
that we measured.  From the ratio of this integral to the
integral made only to the limiting absolute magnitude, we 
obtained correction factors of 1.09 ($z=0.65-0.95$) and 1.24 
($z=0.95-1.15$).  We applied these corrections to the 
measured 840~\AA\ VEMs, giving incompleteness corrected values 
that are not substantially higher than the direct measurements
(horizontal lines in Figure~\ref{vem_cum}).
Numerically these values are
$1.07 \pm 0.19\times 10^{24}$~erg~Mpc$^{-3}$~s$^{-1}$
at $z=0.8$ and 
$1.68\pm0.23\times10^{24}$~erg~Mpc$^{-3}$~s$^{-1}$ 
at $z=1.05$.

\begin{figure}
\hskip -0.2cm
\includegraphics[width=3.6in,clip,angle=0,scale=1]{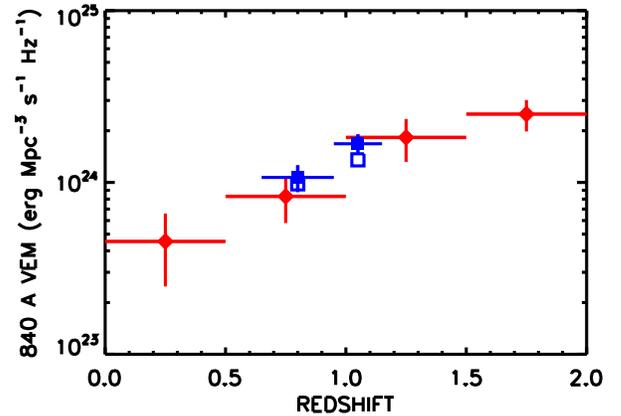}
\caption{Evolution of the comoving volume emissivity
at rest-frame 840~\AA\ vs. redshift for the redshift intervals 
$z=0.65-0.95$ and $z=0.95-1.15$ (blue open squares show the directly 
measured values, and solid squares show the incompleteness 
corrected values).  We also show the evolution determined by 
CBT09 from measuring the {\em GALEX\/} FUV magnitudes of an 
X-ray selected sample (red diamonds).
For each point the $x$-axis error bars show the redshift 
interval and the $y$-axis error bars show the $1\sigma$ 
error computed using the jackknife method.
\label{ion_evol}
}
\end{figure}

We can compare these results to CBT09's determination
of the redshift evolution of the VEM just below the Lyman 
continuum edge.  In Figure~\ref{ion_evol} we show CBT09's 
X-ray based points with red diamonds and the present points 
with blue squares (open for directly measured values; solid 
for incompleteness corrected values).  
The $1\sigma$ errors for both samples have been 
computed with the jackknife method.  The smaller error bars 
on the present points reflect the much larger area 
sampled by the {\em GALEX\/} images than by the X-ray images.  
It is very reassuring that the two samples agree within the 
formal errors.

%------------------------------------------------------------
% FIGURE 5. Ionization evolution: ion_evol.pro
%-------------------------------------------------------------
%--------------------------------------------------------------

%--------------------------------------------
%   6.  The Shape of the Spectrum
%--------------------------------------------
\section{The Shape of the Spectrum} 
\label{secshape}

With the grism data we are able to move one critical step 
beyond CBT09 and compute the shape of the VEM with wavelength. 
As discussed in Section~\ref{secintro}, a direct 
measurement of the shape of the input ionizing source
spectrum is ideally required for the cosmological modeling of 
galaxies and the intergalactic gas.  
For each rest-frame 
wavelength interval we computed the VEM using the same 
procedure that we used for the rest-frame $800-880$~\AA\ 
interval, including applying at all wavelengths the 
incompleteness corrections of 1.09 and 1.24, respectively, 
for the low and high-redshift intervals.  However,
in order to obtain the true rest-frame VEM spectrum,
we also need to correct for the effects of the Lyman scattering 
and absorption in the intervening intergalactic medium 
(M$\o$ller \& Jakobsen 1990; Haardt \& Madau 1996;
Fardal et al.\ 1998).  While this correction is not large 
at $z=1$, it does slightly steepen the spectral slope in the 
extreme UV (EUV) and raise the normalization.

The effects of the intergalactic medium (IGM) may be separated
into scattering by the Ly$\alpha$ forest lines below a rest-frame of 
1216~\AA\ and photoelectric absorption from the Lyman continuum 
of hydrogen below 912~\AA.  At the wavelengths of interest we may 
neglect helium.  The effects of the forest are dominated by systems
with column densities around $10^{14}$~cm$^{-2}$, where Ly$\alpha$ 
lines saturate, while the photoelectric absorption is dominated
by systems with column densities $\sim10^{17}$~cm$^{-2}$,
where the Lyman continuum edge saturates.

The decrement in the Ly$\alpha$-only portion of the forest
has been directly measured by Kirkman et al.\ (2007). They
obtained a value of $0.037\times(1+z)^{1.1}$ at these
redshifts. The redshift dependence, which causes less light
to be absorbed at shorter wavelengths (i.e., because there
are fewer systems at lower redshifts), 
is roughly balanced by the onset of absorption by the higher 
Lyman series lines at shorter wavelengths, resulting in a 
roughly constant reduction as a function of wavelength 
near $z=1$ (see Figure~6a of Inoue \& Iwata 2008).

The photoelectric absorption is most directly related to
the density of Lyman limit systems (LLSs). In LLSs the
optical depth, $\tau$, at the Lyman continuum edge exceeds one.  
The surface density of LLSs at these redshifts has recently been 
remeasured by Songaila \& Cowie (2010).  They found that the number 
density per unit redshift evolves as $0.15\times(1+z)^{1.9}$.
This normalization is slightly lower than previous measurements 
at these redshifts (Stengler-Larrea et al.\ 1995).
We may use this normalization in conjunction with the distribution
of the number of systems as a function of the neutral hydrogen
column density to compute the complete photoelectric effect.
The column density distribution is usually parameterized as a 
power law or broken power law function of the column density. 
Inoue \& Iwata (2008) assumed a broken power law with
a slope of $-1.6$ below the Lyman limit and $-1.3$ above
(e.g., Janknecht et al.\ 2006; Songaila \& Cowie 2010),
which we adopt here.  We have also computed the correction
for a single power law with a slope of $-1.4$ in the vicinity of the
Lyman limit (Misawa et al.\ 2007) to investigate the sensitivity
to the assumed column density distribution. This slightly
decreases the correction, but the changes produced
are smaller than the statistical errors for both the normalization
and the spectral slope.

\begin{figure}
\includegraphics[width=3.6in]{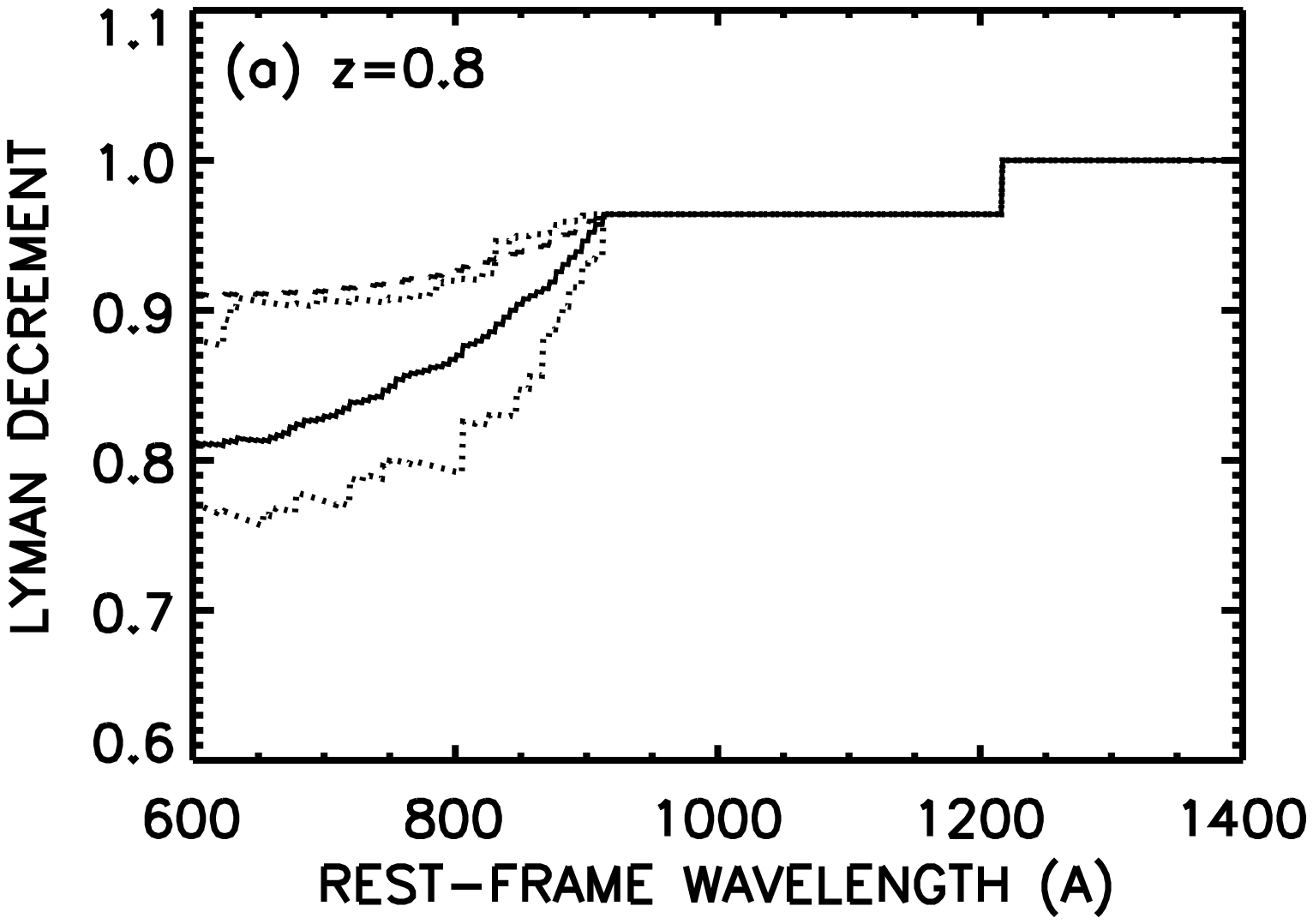}
\includegraphics[width=3.6in]{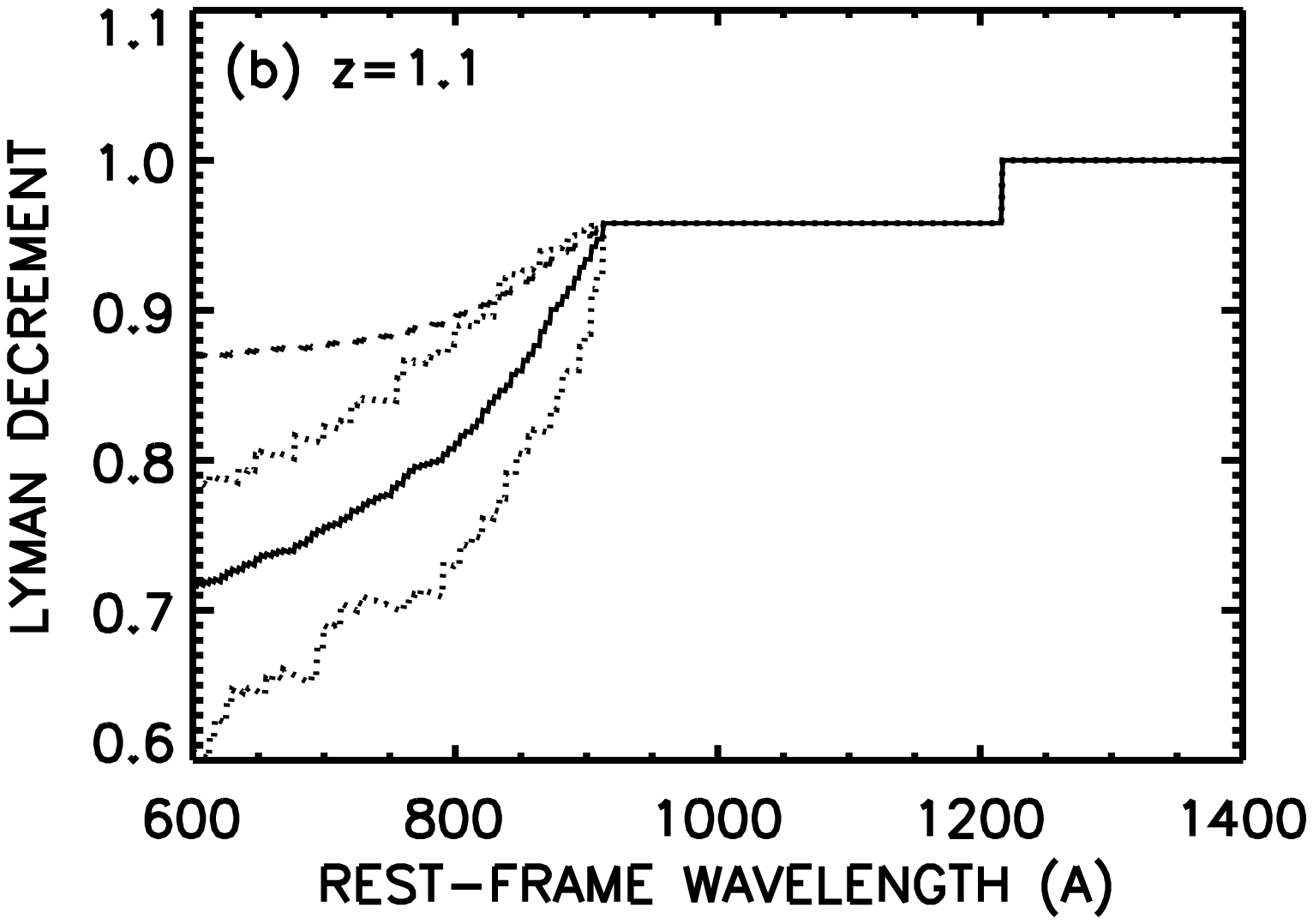}
\caption{
Monte Carlo estimates of the photoelectric absorption 
decrements at (a) $z=0.8$ and (b) $z=1.1$ obtained from
averaging (in each case) 1000 lines of sight.
The dashed curves only show the values from systems with 
optical depths less than one at the Lyman edge, while
the solid curves show the values from all systems regardless 
of their optical depths.  The dotted curves show the range of
corrections for 20 computations of only 60 lines of sight, which
is similar to the number of systems that we have in each redshift 
interval.
\label{figlls}
}
\end{figure}

We have used these values to make Monte Carlo estimates
of the photoelectric absorption (Bershady et al.\ 1999;
Inoue \& Iwata 2008).  In Figure~\ref{figlls} we show 
the photoelectric absorption decrements that we obtained at 
(a) $z=0.8$ and (b) $z=1.1$ from 
averaging (in each case) 1000 lines of sight.
We show the values obtained both from using only
systems with optical depths less than one at the Lyman
edge (dashed line) and from using all the systems 
regardless of their optical depths (solid line). 
The Monte Carlo simulations also allow 
us to estimate the uncertainty introduced by the finite number 
of systems we are observing.  
For the case where we used all the systems regardless of their
optical depth, we also show with dotted 
lines the range of corrections that we found in 20 computations 
using only 60 lines of sight (which is roughly equal to the number 
of lines of sight in the present sample; we have
53 systems at $z=0.65-0.95$ and 63 at $z=0.95-1.25$).
At the shorter wavelengths this corresponds to about a 10\% 
uncertainty in the correction. The large range is a consequence 
of the small number of systems which contribute to the
opacity at $\tau>1$. The uncertainty in the correction for
the $\tau<1$ systems only is much smaller (about 3\% or less).
This is because there is a much larger number of low column density 
systems, and hence the correction is better averaged even in 
a small number of lines of sight.

\begin{figure}
\includegraphics[width=3.6in]{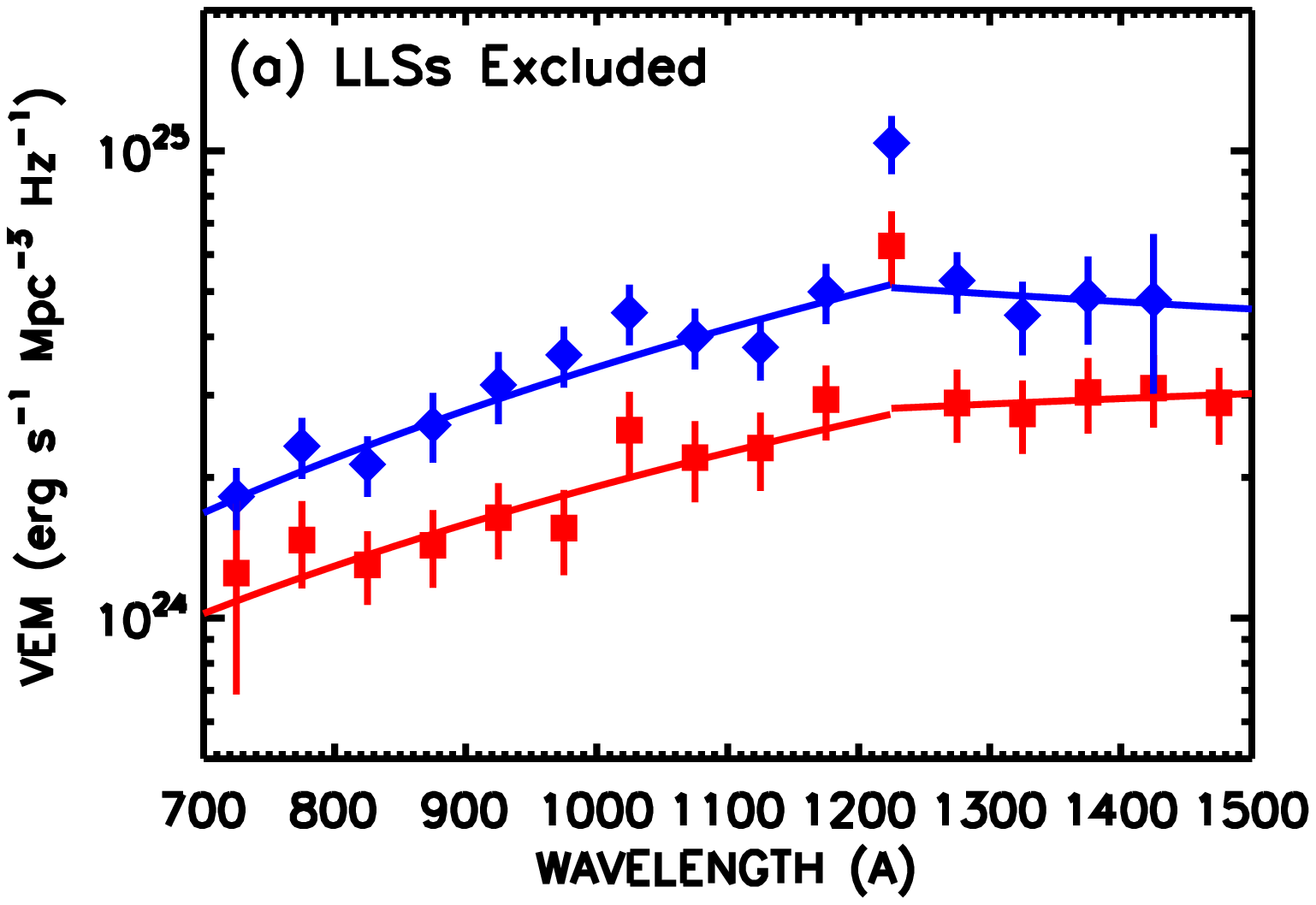}
\includegraphics[width=3.6in]{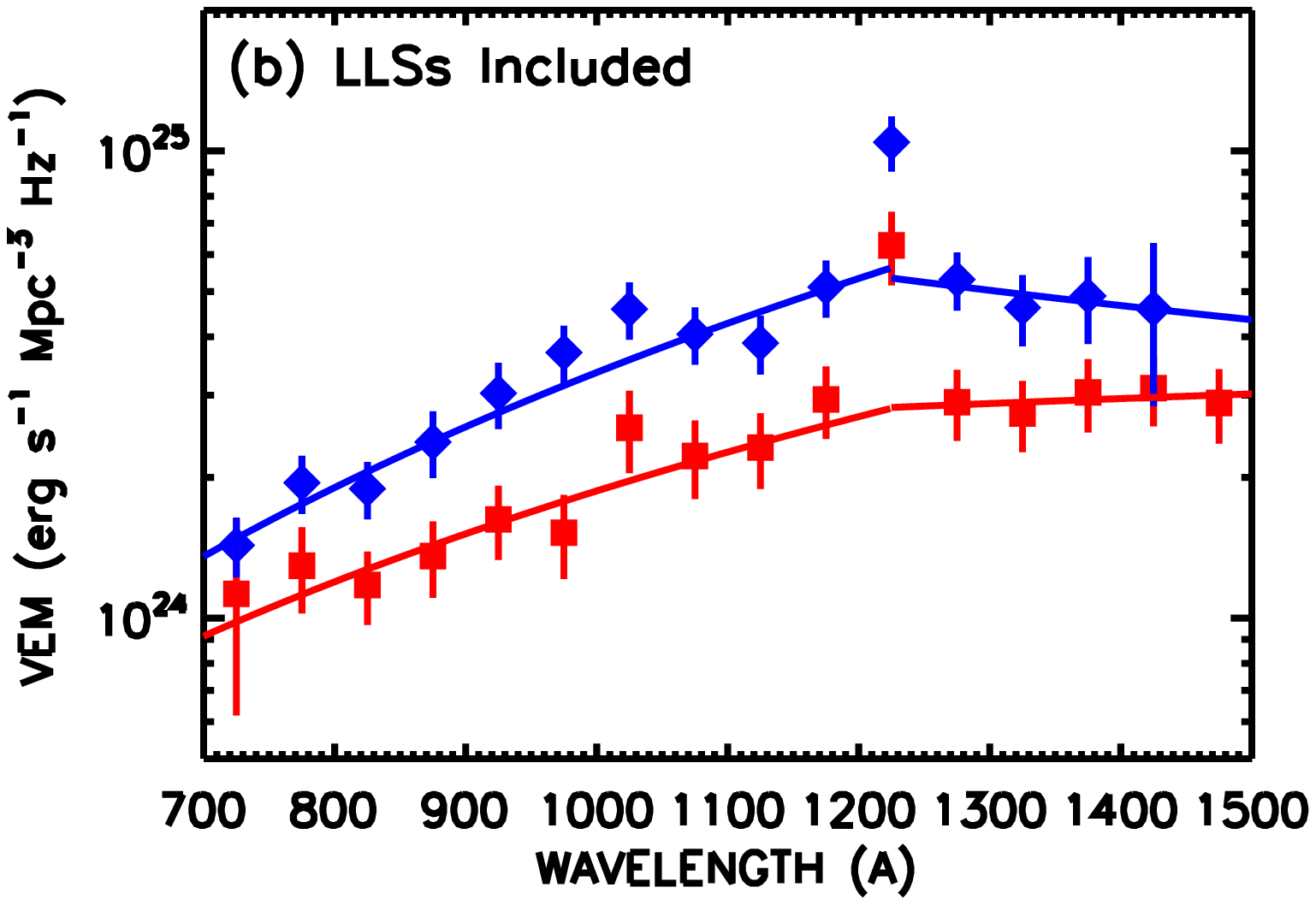}
\caption{Shape of the comoving volume emissivity vs.
rest-frame wavelength for the redshift intervals 
$z=0.65-0.95$ (red squares) and
$z=0.95-1.25$ (blue diamonds). 
The values shown are corrected for incompleteness, and 
the error bars are $\pm1~\sigma$ computed using the jackknife 
method.  
In (a) we exclude the quasars with measured LLSs in their
spectra before applying the $\tau<1$ correction to
compute the final VEM.
In (b) we use the full quasar sample and apply the
full correction, including the opacity from the $\tau>1$
systems, to compute the final VEM.
The straight lines show broken power law fits.
\label{ion_shape}
}
\end{figure}

We have applied the correction in two ways. In the first
we excluded the small number of quasars with measured LLSs 
in their spectra using the list in Songaila \& Cowie (2010).
We corrected for the missing light from these AGNs 
by proportionally reducing the observed area.
We then used the $\tau<1$ correction to compute the final VEM. 
In Figure~\ref{ion_shape}(a) we show the VEM corrected in 
this way and its $1\sigma$ errors in the redshift intervals 
$z=0.65-0.95$ (red squares) and $z=0.95-1.25$ (blue diamonds) 
versus wavelength. The advantage of this method is that it 
minimizes the correction and also avoids the statistical 
uncertainties in the $\tau>1$ correction.  However, we may 
fail to identify LLSs in the fainter and noisier spectra. 
We have therefore also computed the VEM from the full quasar 
sample (i.e., not excluding systems with LLSs).  In this
case we apply the full correction, including the opacity from  
the $\tau>1$ systems.  In Figure~\ref{ion_shape}(b) we show
the VEM corrected in this way and its $1\sigma$ errors in the
same redshift intervals.  We summarize the values that we
derived using both methods in Table~2.  

Scott et al.\ (2004) fitted a single power law to their composite
EUV QSO spectrum from {\em Far-Ultraviolet Spectroscopic Explorer
(FUSE)\/} data, but 
both Zheng et al.\ (1997) and Telfer et al.\ (2002) used a 
broken power law to fit their composite EUV QSO spectra from
{\em Hubble Space Telescope (HST)\/} data.
For each redshift interval we have fitted a broken power
law to the data, excluding the bins at 775~\AA, 1025~\AA, and 
1225~\AA\ to exclude the strong emission lines in the spectra. 
The position of the break is quite uncertain.  We have chosen 
to position it at 1225~\AA.  We show the fits as straight lines
in Figure~\ref{ion_shape} for each case.
Although it is possible that there is more structure
within the error bars, the present data do not justify a more 
complicated fit.  We present this simple parameterization
in order to provide quantitative numbers for use by others.
Since the differences between the two methods lie within the 
statistical errors (see Table~2), we consider only the cases 
with the LLSs excluded (Figure~\ref{ion_shape}a), which should 
be more robust.  For $z=0.65-0.95$ we obtain
\begin{equation}
\log({\rm VEM}) = (24.11\pm0.03) + (1.76\pm0.25)\log(w/800~{\rm \AA}) 
\label{fit1}
\end{equation}
below 1225~\AA\ and
\begin{equation}
\log({\rm VEM}) = (24.46\pm0.01) + (0.36\pm0.45)\log(w/1250~{\rm \AA}) 
\label{fit2}
\end{equation}
above 1225~\AA.  For $z=0.95-1.25$ we obtain
\begin{equation}
\log({\rm VEM}) = (24.34\pm0.02) + (2.01\pm0.20)\log(w/800~{\rm \AA}) 
\label{fit3}
\end{equation}
below 1225~\AA\ and
\begin{equation}
\log({\rm VEM}) = (24.69\pm0.02) - (0.51\pm0.97)\log(w/1250~{\rm \AA}) 
\label{fit4}
\end{equation}
above 1225~\AA.  In all of the above equations, $w$ is the rest-frame 
wavelength.

The slopes are consistent between the two redshift ranges within
the statistical errors, while the normalization rises by 0.23~dex 
or a factor of 1.7.  The short wavelength slope is considerably
steeper than the value of 0.56 $(-0.28,0.38)$ derived by 
Scott et al.\ (2004) but agrees well with 
the values of $1.76\pm0.12$ derived by Telfer et al.\ (2002) and 
$1.96\pm0.15$ derived by Zheng et al.\ (1997).
These various samples have different luminosity and redshift
selections, which may account for some part of the differences.
The present sample is properly weighted over the sources contributing 
to the ionizing background.

%------------------------------------------------------------
% FIGURE 6.  lls_dec_a.pro, lls_dec_b.pro
%-------------------------------------------------------------
%-------------------------------------------------------------

%----------------------------------------------------------------
% FIGURE 7. Shape of ionizing source function (lam_ion_shape.pro)
%----------------------------------------------------------------
%--------------------------------------------------------------

%---------------
%  Summary
%---------------
\section{Summary}
\label{secsum}

We used a complete sample of Ly$\alpha$ emission-line
selected AGNs obtained from nine deep blank fields observed
with the grism spectrographs on {\em GALEX\/} to measure
the AGN contribution to the ionizing background at $z\sim 1$.
We found that the {\em GALEX\/} AGN luminosity function is
in good agreement with luminosity functions obtained 
from optical and X-ray AGN samples.  We also found that
our measurements of the 840~\AA\ VEM at $z=0.8$ and $z=1.05$ 
agree within the formal errors with the 
redshift evolution of the comoving volume emissivity 
determined by CBT09 using an X-ray selected AGN sample.
Finally, for the first time, we were able 
to compute directly the shape of the VEM with wavelength 
for the redshift intervals $z=0.65-0.95$ and $z=0.95-1.25$.
We corrected both for incompleteness and for the effects
of the Lyman scattering and absorption in the intervening
intergalactic medium.  The slopes are consistent between
the two redshift ranges within the statistical errors,
while the normalization rises by a factor of 1.7.
At $<1125$~\AA\ our slope is considerably steeper than
that determined by Scott et al.\ (2004) using {\em FUSE\/}
data, but it is in good agreement with the slopes determined 
by Telfer et al.\ (2002) and Zheng et al.\ (1997) using
{\em HST\/} data.

%\vskip 20in
\acknowledgements

We thank the anonymous referee for a helpful report.
We gratefully acknowledge support from NSF grants 
AST-0708793 (A.~J.~B.) and AST-0709356 (L.~L.~C.),
the University of Wisconsin Research Committee with 
funds granted by the Wisconsin Alumni Research Foundation,
and the David and Lucile Packard Foundation (A.~J.~B.).

%\newpage

%------------------
%   References
%-------------------

%-------------
% TABLE 2
%-------------
\begin{deluxetable}{cccccc}
\tablewidth{0pt}
\tablecaption{VEM}
\scriptsize
\tablehead{Wavelength (\AA) & Raw VEM\tablenotemark{a} & Error\tablenotemark{a} & Lyman Corr. & Final VEM & Error \\ 
(1) & (2) & (3) & (4) & (5) & (6)
}
\startdata
\multicolumn{6}{c}{LLSs Excluded, $z=0.65-0.95$} \cr
\hline
725 &  0.96  &  0.43  &  0.92  & 1.13 &  0.51 \cr
775  & 1.11  &  0.23  &  0.92  & 1.29 &  0.27 \cr
825  & 1.03   & 0.18  &  0.93  & 1.18 &  0.21 \cr
875   & 1.20  & 0.22   & 0.95 &  1.36  & 0.25 \cr
925   & 1.47  & 0.26   & 0.96 &  1.63  & 0.29 \cr
975   & 1.37   & 0.28  &  0.96  & 1.52 &  0.31 \cr
1025   & 2.30  & 0.46   & 0.96  &  2.56 &  0.51 \cr
1075   & 2.00  & 0.38  & 0.96 &  2.22  & 0.42 \cr
1125   & 2.09  & 0.39  & 0.96 &  2.32  & 0.43 \cr
1175   & 2.65 &  0.47  & 0.96 &  2.94  & 0.52 \cr
1225   & 5.87  & 1.05   & 1.00 &  6.28  & 1.13  \cr
1275   & 2.71  & 0.47 &  1.00 &  2.90 &  0.50  \cr
1325  &  2.56  & 0.44  & 1.00  & 2.74 &  0.48  \cr
1375   & 2.84  & 0.51  & 1.00  & 3.04  & 0.54  \cr
\hline 
\multicolumn{6}{c}{LLSs Excluded, $z=0.95-1.25$} \cr
\hline
725   & 1.01   & 0.15  & 0.88 &  1.43   & 0.21 \cr
775   &  1.39  & 0.20  & 0.89 &  1.95  &  0.28 \cr
825   &  1.37  & 0.19  & 0.90  & 1.89 &  0.27 \cr
875   &  1.78  & 0.29  & 0.93 &  2.38  & 0.39 \cr
925  &  2.32  & 0.38  & 0.96  & 3.03  & 0.49 \cr
975   & 2.84  & 0.40 &  0.96 &  3.70 &  0.52 \cr
1025  &  3.51  & 0.49  & 0.96  & 4.59 &  0.64 \cr
1075   & 3.11 &  0.44  & 0.96  & 4.05 &  0.57 \cr
1125  & 2.97  & 0.43  & 0.96  & 3.88 &  0.56 \cr
1175  & 3.92  & 0.55  & 0.96  & 5.11  & 0.72 \cr
1225  &  8.35  & 1.13  & 1.00 & 10.44 &  1.41 \cr
1275  &  4.24  & 0.60  & 1.00  & 5.31 &  0.76 \cr
1325  &  3.70  & 0.64  & 1.00  & 4.62  & 0.80 \cr
1375   & 3.92 &  0.82   & 1.00 &  4.89  & 1.03 \cr
%\tablebreak
\hline
\multicolumn{6}{c}{LLSs Included, $z=0.65-0.95$} \cr
\hline
725   & 0.98 &  0.44 &  0.84  & 1.25  & 0.56 \cr
775  & 1.18  & 0.25  & 0.86  & 1.47 &  0.31  \cr
825 &  1.07 &   0.19  & 0.88 &  1.30 &   0.23   \cr
875 &  1.23  & 0.23   & 0.92 &  1.43 &  0.27   \cr
925  & 1.48  & 0.27 &  0.96 &  1.64 &  0.30    \cr
975  & 1.40  & 0.29 &  0.96 &  1.56 &  0.32     \cr
1025  & 2.27 &  0.47 &  0.96 &  2.52 &  0.53   \cr
1075  &  1.99  & 0.39 &  0.96 &  2.20 &  0.44     \cr
1125 &  2.08  & 0.40 &  0.96 &  2.31 &  0.44    \cr
1175  & 2.64  & 0.48 &  0.96 &  2.94 &  0.53     \cr
1225  & 5.85  & 1.08 &  1.00 &   6.26 &  1.16    \cr
1275  & 2.70 &  0.48 &  1.00  &  2.89 &  0.51   \cr
1325  & 2.56  & 0.46 &  1.00  & 2.74 &  0.49   \cr
1375  & 2.84  & 0.52 &  1.00  & 3.04 &  0.56    \cr
\hline
\multicolumn{6}{c}{LLSs Included, $z=0.95-1.25$} \cr
\hline \cr
725 &  1.11 &  0.17 &  0.77   & 1.82 &  0.28 \cr
775 & 1.49 &  0.22  & 0.80 &  2.33 &  0.35 \cr
825  &  1.42  & 0.21  & 0.83 &  2.13 &  0.32 \cr
875  &  1.87 &  0.32 &  0.90 &   2.59 &  0.44 \cr
925  &   2.42 &  0.42 &  0.96 & 3.16 &  0.55 \cr
975 &   2.80  & 0.42 &  0.96 &  3.66 &  0.55 \cr
1025  & 3.45  & 0.51  & 0.96 &  4.50 &  0.67 \cr
1075  &  3.07  & 0.45 &  0.96  & 4.00 &  0.59 \cr
1125  &  2.91  & 0.44  & 0.96 &  3.80 &  0.57 \cr
1175  &  3.83 &  0.56  & 0.96 &  4.99  & 0.73 \cr
1225  &  8.32  & 1.18  & 1.00 & 10.39 &  1.48 \cr
1275  &  4.22  & 0.64 &  1.00 &  5.28 &  0.79 \cr
1325  &  3.56  & 0.64  & 1.00  & 4.45 &  0.80 \cr
1375  &  3.92  & 0.84 &  1.00  & 4.89 & 1.04 \cr
\enddata
\label{tab2}
\tablenotetext{a}{Raw values do not have the incompleteness
correction or the Lyman correction.  The VEM and errors are
given in units of $10^{24}$~erg~Mpc$^{-3}$~s$^{-1}$~Hz$^{-1}$.
}
\end{deluxetable}

\end{document}